\documentclass[12pt, centerh1]{article}
\usepackage{natbib}
\usepackage[margin=1in]{geometry}
\usepackage{amsmath,natbib,multirow}
\usepackage{amsfonts,mathrsfs,moreverb,lipsum}
\usepackage{graphicx,colonequals}
\usepackage{fixltx2e}
\usepackage{color}
\usepackage{caption}
\usepackage{subcaption}
\usepackage{mathtools}
\usepackage{pbox}
\newcommand{\btheta}{{\boldsymbol{\theta}}}
\newcommand{\bvtheta}{{\boldsymbol{\vartheta}}}
\newcommand{\pig}{\pi_g}

\newcommand{\vecmu}{\mbox{\boldmath$\mu$}}

\newcommand{\vecx}{\mathbf{x}}

\newcommand{\vecX}{\mathbf{X}}
\newcommand{\vecU}{\mathbf{U}}

\newcommand{\vecW}{\mathbf{W}}

\newcommand{\vecz}{\mathbf{z}}

\newcommand{\loada}{\mathbf\Lambda}
\newcommand{\loadb}{\mathbf\Delta}
\newcommand{\matsig}{\mathbf\Sigma}
\newcommand{\matPsi}{\mathbf\Psi}

\newcommand{\diag}{\,\mbox{diag}}
\newcommand{\tr}{\,\mbox{tr}}

\newcommand{\vecB}{\mathbf{B}}

\newcommand{\vecS}{\mathbf{S}}
\newcommand{\vecP}{\mathbf{P}}

\newcommand{\matm}{\mathbf{M}}

\newcommand{\ident}{\mathbf{I}}

\newcommand{\vecepsilon}{\mbox{\boldmath$\varepsilon$}}

\newcommand{\vecc}{\text{vec}}
\newcommand{\fX}{\mathscr{X}}
\newcommand{\fV}{\mathscr{V}}
\newcommand{\fU}{\mathscr{U}}
\newcommand{\fE}{\mathscr{E}}
\newcommand{\fY}{\mathscr{Y}}

\title{Parsimonious Mixtures of Matrix Variate Bilinear Factor Analyzers}
\author{Michael P.B. Gallaugher and Paul D. McNicholas}
\date{\small Dept.\ of Mathematics \& Statistics, McMaster University, Hamilton, Ontario, Canada.}
\begin{document}
\maketitle{}
\begin{abstract}
Over the years, data have become increasingly higher dimensional, which has prompted an increased need for dimension reduction techniques. This is perhaps especially true for clustering (unsupervised classification) as well as semi-supervised and supervised classification. Many methods have been proposed in the literature for two-way (multivariate) data and quite recently methods have been presented for three-way (matrix variate) data. One such such method is the mixtures of matrix variate bilinear factor analyzers (MMVBFA) model. Herein, we propose of total of 64 parsimonious MMVBFA models. Simulated and real data are used for illustration.\\[-10pt]

\end{abstract}

\section{Introduction}
One aspect of more complex data collected today is the increasing dimensionality of the data. Therefore, there has been an increased need for parameter reduction and dimension reduction techniques, especially in the area of model-based clustering. Such techniques are abundant in the literature for traditional, multivariate data and include parsimonious models \citep{celeux95}, mixtures of factor analyzers \citep{ghahramani97, mcnicholas08}, co-clustering \citep{hartigan72, nadif10, gallaugher18c}, and penalization methods \citep{pan07,zhou09pen,gallaugher19b}. In the case of three-way data, however, there are still considerable gaps in the literature for clustering high-dimensional three-way data.

Three-way data comes in the form of matrices, examples of which include greyscale images and multivariate longitudinal data --- the latter consists of multiple variables collected at different time points. In the last few years, many methods have been proposed for analyzing three-way data. One recent method is the mixture of matrix variate bilinear factor analyzers model \citep{gallaugher18b}, which can be considered the matrix variate analogue of the mixture of factor analyzers model. Herein, we present a total of 64 parsimonious models in the matrix variate case, which is effectively a matrix variate analogue of the multivariate family presented by \cite{mcnicholas08}. The remainder of this paper is laid out as follows. In Section~2, some background on model-based clustering and matrix variate methods is presented. In Section~3, the methodology is outlined. Then, simulations and data analyses are presented in Sections~4 and 5, respectively. We conclude with a discussion and suggestions for future work (Section~6).

\section{Background}
\subsection{Model-Based Clustering}
Clustering is the process of finding homogenous group structure within heterogenous data. One of the most established methods in the literature is model-based clustering, which makes use of a finite mixture model. A finite mixture model assumes that a random variable $\vecX$ comes from a population with $G$ subgroups and its density can be written 
$$
f(\vecx~|~\bvtheta)=\sum_{g=1}^G\pi_g f_{g}(\vecx~|~\btheta_g),
$$
where $\sum_{g=1}^G\pi_g=1$, with $\pi_g>0$, are the mixing proportions and $f_g(\cdot)$ are the component densities.
The mixture of Gaussian distributions has been studied extensively within the literature, with \cite{wolfe65} being an early example of the use of a mixture of Gaussian distributions for clustering. Other early examples of clustering with a mixture of Gaussians can be found in \cite{baum70} and \cite{scott71}.

Because of the flexibility of the mixture modelling framework, many other mixtures have been proposed using more flexible distributions such as those that allow for parameterization of tail weight such as the $t$ distribution \citep{peel00,andrews11a,andrews12,lin14} and the power exponential distribution \citep{dang15}, as well as those that allow for the parameterization of skewness and tail weight such as the skew-$t$ distribution \citep{lin10,vrbik12,vrbik14,lee14,murray14b,murray14a} and others \citep{browne15,franczak15,murray17b,tang18,tortora19}. 

In addition to the multivariate case, there are recent examples of using matrix variate distributions for clustering three-way data. Such examples include using the matrix variate normal \citep{viroli11}, skewed distributions \citep{gallaugher17a,gallaugher18a, gallaugher19}, and transformation methods \citep{melnykov18}. Most recently, \cite{sarkar20} present parsimonious models analogous to those used by \cite{celeux95}.

\subsection{Parsimonious Gaussian Mixture Models}
One popular dimension reduction technique for high dimensional multivariate data is the mixture of factor analyzers model. The factor analysis model for $p$-dimensional $\vecX_1,\ldots,\vecX_N$ is given by
$$
\vecX_i=\vecmu+\loada\vecU_i+\vecepsilon_i,
$$
where $\vecmu$ is a location vector, $\loada$ is a $p\times q$ matrix of factor loadings with $q<p$, $\vecU_i\sim\mathcal{N}_q({\bf 0},\ident)$ denotes the latent factors, $\vecepsilon_i\sim \mathcal{N}_q({\bf 0},\matPsi)$, where $\matPsi=\text{diag}(\psi_1,\psi_2,\ldots,\psi_p)$, $\psi_j\in \mathbb{R}^+$, and $\vecU_i$ and $\vecepsilon_i$ are each independently distributed and independent of one another. Under this model, the marginal distribution of $\vecX_i$ is $\mathcal{N}_p(\vecmu,\loada\loada'+\matPsi)$. Probabilistic principal component analysis (PPCA) arises as a special case with the isotropic constraint $\matPsi=\psi\ident$, $\psi\in\mathbb{R}^+$ \citep{tipping99a}. 

\cite{ghahramani97} develop the mixture of factor analyzers model, where the density takes the form of a Gaussian mixture model with covariance structure $\matsig_g=\loada_g\loada_g'+\matPsi$. A small extension was presented by \cite{mclachlan00a}, who utilize the more general structure $\matsig_g=\loada_g\loada_g'+\matPsi_g$. \cite{tipping99b} introduce the closely-related mixture of PPCAs with $\matsig_g=\loada_g\loada_g'+\psi_g\ident$. \cite{mcnicholas08} consider all combinations of the constraints $\loada_g=\loada$, $\matPsi_g=\matPsi$, and the isotropic constraint to give a family of eight parsimonious Gaussian mixture models (PGMMs). As discussed by \cite{mcnicholas08}, the number of covariance parameters for each PGMM is linear in the data dimension $p$ as compared to the parsimonious models presented by \cite{celeux95}, where the majority are quadratic in $p$ and the others assume variable independence. This paper introduces a matrix variate analogue of the PGMM family of \cite{mcnicholas08}.

\subsection{Matrix Variate Normal Distribution}
In recent years, several methods have been proposed for clustering three-way data. These methods mainly employ the use of matrix variate distributions in finite mixture models. Like the univariate and multivariate cases, the most mathematically tractable matrix variate distribution to use is the matrix variate normal distribution. An $n\times p$ random matrix $\fX$ follows a matrix variate normal distribution with location parameter $\matm$ and scale matrices $\matsig$ and $\matPsi$ of dimensions $n\times n$ and $p\times p$, respectively, denoted by $\mathcal{N}_{n\times p}(\matm, \matsig, \matPsi)$, if the density of $\fX$ can be written
\begin{equation}
f(\vecX~|~\matm, \matsig, \matPsi)=\frac{1}{(2\pi)^{\frac{np}{2}}|\matsig|^{\frac{p}{2}}|\matPsi|^{\frac{n}{2}}}\exp\left\{-\frac{1}{2}\tr\left(\matsig^{-1}(\vecX-\matm)\matPsi^{-1}(\vecX-\matm)'\right)\right\}.
\end{equation}
One notable property of the matrix variate normal distribution \citep{harrar08} is
\begin{equation}
\fX\sim \mathcal{N}_{n\times p}(\matm,\matsig,\matPsi) \iff \vecc(\fX)\sim \mathcal{N}_{np}(\vecc(\matm),\matPsi\otimes \matsig),
\label{eq:normprop}
\end{equation}
where $\mathcal{N}_{np}(\cdot)$ is the $np$-dimensional multivariate normal density, $\vecc(\cdot)$ is the vectorization operator, and $\otimes$ denotes the Kronecker product. 

\subsection{Mixture of Matrix Variate Bilinear Factor Analyzers} 
\cite{gallaugher18b} present an extension of the work of \cite{xie08}, \cite{yu08} and \cite{zhao12} to derive the MMVBFA model. The MMVBFA model assumes that
\begin{equation}
\fX_i=\matm_g+\loada_g\fU_{ig}\loadb_g'+\loada_g\fE_{ig}^B+\fE^A_{ig}\loadb_g'+\fE_{ig}
\label{eq:FAmod}
\end{equation}
with probability $\pi_g$, for $g=1,2,\ldots,G$, where $\matm_g$ is an $n\times p$ location matrix, $\loada_g$ is an $n\times q$ column factor loading matrix, with $q<n$, $\loadb_g$ is a $p\times r$ row factor loading matrix, with $r<p$, and
\begin{equation*}\begin{split}
\fU_{ig}&\sim \mathcal{N}_{q\times r}({\bf 0},\ident_q,\ident_r),\\
\fE_{ig}^{B}&\sim  \mathcal{N}_{q\times p}({\bf 0},\ident_q,\matPsi_g),\\
\fE_{ig}^{A}&\sim  \mathcal{N}_{n\times r}({\bf 0},\matsig_g,\ident_r),\\
\fE_{ig}&\sim   \mathcal{N}_{n\times p}({\bf 0},\matsig_g,\matPsi_g)
\end{split}\end{equation*}
are independent of each other, $\matsig_g=\diag\{\sigma_{g1},\sigma_{g2},\ldots,\sigma_{gn}\}$, with $\sigma_{gj}\in \mathbb{R}^+$, and $\matPsi_g=\diag\{\psi_{g1},\psi_{g2}, \ldots, \psi_{gp}\}$, with $\psi_{gj}\in\mathbb{R}^+$. Let $\vecz_i=(z_{i1},\ldots,z_{iG})'$ denote the component membership for $\vecX_i$, where
$$ 
z_{ig}=
\begin{cases}
1 & \mbox{if } \vecX_i \mbox{ belongs to component } g,\\
0 &\mbox{otherwise},
\end{cases}
$$ for $i=1,\ldots,N$ and $g=1,\ldots,G$. Using the vectorization of $\fX_i$, and property \eqref{eq:normprop}, it can be shown that 
$$
\fX_i~|~z_{ig}=1 \sim \mathcal{N}_{n\times p}(\matm_g,\matsig_g^{*},\matPsi_g^{*}),
$$
where $\matsig_g^{*}=\matsig_g+\loada_g\loada_g'$ and $\matPsi_g^{*}=\matPsi_g+\loadb_g\loadb_g'$.
Therefore, the density of $\fX_i$ can be written
$$
f(\vecX_i~|~\bvtheta)=\sum_{g=1}^G\pig \varphi_{n\times p}(\vecX_i~|~\matm_g,\matsig_g^{*},\matPsi_g^{*}),
$$
where $\varphi_{n\times p}(\cdot)$ denotes the $n\times p$ matrix variate normal density.

Note that the term ``column factors'' refers to reduction in the dimension of the columns, which is equivalent to the number of rows, and not a reduction in the number of columns. Likewise, the term ``row factors'' refers to the reduction in the dimension of the rows (number of columns). Moreover, as discussed by \cite{zhao12}, we can interpret terms $\fE^B$ and $\fE^A$ as the row and column noise, respectively, and the final term $\fE$ as the common noise. 

As discussed by \cite{zhao12} and \cite{gallaugher18a}, by introducing latent variables $\fY_{ig}^R$ and $\fV_{ig}^R$, \eqref{eq:FAmod} exhibits the two-stage formulation 
\begin{equation*}\begin{split}
\fX_i&=\matm_g+\loada_g\fY_{ig}^B+\fV_{ig}^B,\\
\fY_{ig}^B&=\fU_{ig}\loadb_g'+\fE_{ig}^B,\\
\fV_{ig}^B&=\fE_{ig}^A\loadb_{g}'+\fE_{ig}.
\end{split}\end{equation*}
This formulation can viewed as first projecting $\fX_i$ in the column direction onto the latent matrix $\fY_{ig}^B$, and then $\fY_{ig}^B$ and $\fV_{ig}^B$ are further projected in the row direction.
Likewise, introducing $\fY_{ig}^C$ and $\fV_{ig}^C$, \eqref{eq:FAmod} can be written
\begin{equation*}\begin{split}
\fX_i&=\matm_g+\fY_{ig}^A\loadb_g'+\fV_{ig}^A,\\
\fY_{ig}^A&=\loada_g\fU_{ig}+\fE_{ig}^A,\\
\fV_{ig}^A&=\loada_g\fE_{ig}^B+\fE_{ig}.
\end{split}\end{equation*}
The interpretation is the same as before but we now project in the row direction first followed by the column direction.

\section{Methodology}
\subsection{Parsimonious MMVBFA Models}
One feature of the MMVBFA model is that each of the resultant scale matrices has the same form as the covariance matrix in the (multivariate) mixture of factor analyzers model. Therefore, MMVBFA lends itself naturally to a matrix variate extension of the PGMM models. Specifically, we apply combinations of the constraints $\loada_g=\loada$, $\matsig_g=\matsig$, $\matsig_g=\sigma_g\ident_n$ with $\sigma_g\in\mathbb{R}^+$, $\loadb_g=\loadb$, $\matPsi_g=\matPsi$, and $\matPsi_g=\psi_g\ident_p$ with $\psi_g\in\mathbb{R}^+$. This leads to a total of 64 models, which we refer to as the parsimonious mixtures of matrix variate bilinear factor analyzers (PMMVBFA) family. In Tables \ref{tab:Sigma} and \ref{tab:Psi}, the models along with the number of scale parameters are presented for the row and column scale matrices. We will refer to these as the row and column models, respectively. 
\begin{table}[!t]
\caption{Row models with the respective number of scale parameters.}
\label{tab:Sigma}       
%
%
\begin{tabular*}{1.00\textwidth}{@{\extracolsep{\fill}}p{2cm}p{2.4cm}p{2cm}p{5.9cm}}
\hline
$\loada_g=\loada$&$\matsig_g=\matsig$&$\matsig_g=\sigma_g\ident_n$&Number of Scale Parameters\\
\hline
C&C&C&$[nq+n-q(q-1)/2]+1$\\
C&C&U&$[nq+n-q(q-1)/2]+n$\\
C&U&C&$[nq+n-q(q-1)/2]+G$\\
C&U&U&$[nq+n-q(q-1)/2]+nG$\\
U&C&C&$G[nq+n-q(q-1)/2]+1$\\
U&C&U&$G[nq+n-q(q-1)/2]+n$\\
U&U&C&$G[nq+n-q(q-1)/2]+G$\\
U&U&U&$G[nq+n-q(q-1)/2]+nG$\\
\hline
\end{tabular*}
\end{table}
\begin{table}[!t]
\caption{Column models with the respective number of scale parameters.}
\label{tab:Psi}       
%
%
\begin{tabular*}{1.00\textwidth}{@{\extracolsep{\fill}}p{2cm}p{2.4cm}p{2cm}p{5.9cm}}
\hline
$\loadb_g=\loadb$&$\matPsi_g=\matPsi$&$\matPsi_g=\psi_g\ident_r$&Number of Scale Parameters\\
\hline
C&C&C&$[pr+p-r(r-1)/2]+1$\\
C&C&U&$[pr+p-r(r-1)/2]+p$\\
C&U&C&$[pr+p-r(r-1)/2]+G$\\
C&U&U&$[pr+p-r(r-1)/2]+pG$\\
U&C&C&$G[pr+p-r(r-1)/2]+1$\\
U&C&U&$G[pr+p-r(r-1)/2]+p$\\
U&U&C&$G[pr+p-r(r-1)/2]+G$\\
U&U&U&$G[pr+p-r(r-1)/2]+pG$\\
\hline
\end{tabular*}
\end{table}

Maximum likelihood estimation is performed using an alternating expectation maximization (AECM) algorithm in almost an identical fashion to \cite{gallaugher18b}. The only difference is the form of the updates for the scale matrices which is dependent on the model. Below, the general form of the algorithm is presented and the corresponding scale parameter updates are given in Appendix~A. We refer the reader to \cite{gallaugher18b} for details regarding the expectations in the E-steps.

\paragraph{AECM Stage 1} In the first stage, the complete-data is taken to be the observed matrices $\vecX_1,\ldots,\vecX_N$ and the component memberships $\vecz_1,\ldots,\vecz_N$, and the update for $\matm_g$ is calculated. The complete-data log-likelihood in the first stage is then
$$
\ell^{(1)}=C+\sum_{g=1}^G\sum_{i=1}^N z_{ig}\left\{\log\pig-\frac{1}{2} \tr[\matsig^{*^{-1}}_g(\vecX_i-\matm_g)\matPsi^{*^{-1}}_g(\vecX_i-\matm_g)']\right\},
$$
where $C$ is a constant with respect to $\matm_g$, $\matsig^{*}_g\colonequals\loada_g\loada_g'+\matsig_g$ and $\matPsi^{*}_g\colonequals\loadb_g\loadb_g'+\matPsi_g$.
In the E-Step, the updates for the component memberships $z_{ig}$ are given by the expectations
$$
\hat{z}_{ig}=\frac{\pig\varphi_{n\times p}(\vecX_i~|~\hat{\matm}_g,\hat{\matsig}^{*}_g,\hat{\matPsi}^{*}_g)}{\sum_{h=1}^G\pi_h\varphi_{n\times p}(\vecX_i~|~\hat{\matm}_h,\hat{\matsig}^{*}_h,\hat{\matPsi}^{*}_h)},
$$
where $\varphi_{n\times p}(\cdot)$ denotes the $n\times p$ matrix variate normal density.
In the CM-step, the update for $\matm_g$ is 
$$\hat{\matm}_g=\frac{1}{N_g}\sum_{i=1}^N\hat{z}_{ig}\vecX_{i},$$
where $N_g=\sum_{i=1}^N\hat{z}_{ig}$.

\paragraph{AECM Stage 2} In the second stage, the complete-data is taken to be the observed $\vecX_1,\ldots,\vecX_N$, the component memberships $\vecz_1,\ldots,\vecz_N$ and the latent factors ${\bf \fY}_{i}^B=(\fY_{i1}^B,\fY_{i2}^B,\ldots,\fY_{iG}^B)$. The complete-data log-likelihood is then
\begin{equation}\begin{split}
&\ell^{(2)}=C-\sum_{g=1}^G\frac{N_gp}{2}\log|\matsig_g|-\frac{1}{2}\sum_{g=1}^G\sum_{i=1}^Nz_{ig}\text{tr}\big[\matsig_g^{-1}(\vecX_i-\matm_g)\matPsi^{*^{-1}}_g(\vecX_i-\matm_g)'\\
&\ -\matsig_g^{-1}\loada_g\fY_{ig}^B\matPsi^{*^{-1}}_g(\vecX_i-\matm_g)'-\matsig_{g}^{-1}(\vecX_i-\matm_g)\matPsi^{*^{-1}}_g{\fY_{ig}^B}'\loada_g'+\matsig_g^{-1}\loada_g\fY_{ig}^B\matPsi^{*^{-1}}_g{\fY_{ig}^{B}}'\loada_g'\big].
\end{split}\end{equation}
In the E-Step, the following expectations are calculated:
\begin{equation}\begin{split}
a^B_{ig}&\colonequals\mathbb{E}[\fY_{ig}^B~|~\vecX_{i},z_{ig}=1]={\vecW^{A}_g}^{-1}\loada_g'\matsig_g^{-1}(\vecX_i-\matm_g),\\
b^B_{ig}&\colonequals\mathbb{E}[\fY_{ig}^B\hat{\matPsi}^{*^{-1}}{\fY_{ig}^B}'~|~\vecX_{i},z_{ig}=1]=p{\vecW^{A}_g}^{-1}+a^B_{ig}\hat{\matPsi}^{*^{-1}}_g{a^B_{ig}}',
\end{split}
\end{equation}
where $\vecW^{A}_g=\ident_{q}+\loada_g'\matsig_g^{-1}\loada_g$.
In the CM-step, $\loada_g$ and $\matsig_g$ are updated (see Appendix~A).

\paragraph{AECM Stage 3} In the last stage of the AECM algorithm, the complete-data is taken to be the observed $\vecX_1,\ldots,\vecX_N$, the component memberships $\vecz_1,\ldots,\vecz_N$ and the latent factors ${\bf \fY}_{i}^A=(\fY_{i1}^A,\fY_{i2}^A,\ldots,\fY_{iG}^A)$. In this step, the complete-data log-likelihood is
\begin{equation*}\begin{split}
\ell^{(3)}&=C-\frac{N_g n}{2}\log|\matPsi_g|-\frac{1}{2}\sum_{g=1}^G\sum_{i=1}^N z_{ig}\tr\big[\matPsi_g^{-1}(\vecX_i-\matm_g)'\matsig^{*^{-1}}_g(\vecX_i-\matm_g)\\
&-\matPsi_g^{-1}\loadb_g{\fY_{ig}^A}'\matsig^{*^{-1}}_g(\vecX_i-\matm_g)-\matPsi_g^{-1}(\vecX_i-\matm_g)'\matsig^{*^{-1}}_g{\fY_{ig}^A}\vecB_g'+\matPsi_g^{-1}\loadb_g{\fY_{ig}^A}'\matsig^{*^{-1}}_g{\fY_{ig}^{A}}\loadb_g'\big].
\end{split}\end{equation*}
In the E-Step, expectations similar to those at Stage 2 are calculated:
$$a^A_{ig}:=\mathbb{E}[\fY_{ig}^A~|~\vecX_{i},z_{ig}=1]=(\vecX_i-\matm_g)\matPsi_g^{-1}\loadb_g{\vecW_g^{B}}^{-1}$$
and
$$
b^A_{ig}:=\mathbb{E}[{\fY_{ig}^A}'\hat{\matsig}^{*^{-1}}_g\fY_{ig}^A~|~\vecX_{i},z_{ig}=1]=n{\vecW^{B}_g}^{-1}+{a^A_{ig}}'\hat{\matsig}^{*^{-1}}_g{a^A_{ig}},
$$
where $\vecW^{B}_g=\ident_r+\loadb_{g}'\matPsi_g^{-1}\loadb_g$.
In the CM-step, we update $\loadb_g$ and $\matPsi_g$ (see Appendix~A).

\subsection{Model Selection, Convergence, Performance Evaluation Criteria, and Initialization}
In general, the number of components, row factors, column factors, row model, and column model are unknown {\it a priori} and, therefore, need to be selected. In our simulations and analyses, the Bayesian information criterion \cite[BIC;][]{schwarz78} is used. The BIC is given by
$$
\text{BIC}=2\ell(\hat{\bvtheta})-\rho\log N,
$$
where $\ell(\hat{\bvtheta})$ is the maximized log-likelihood and $\rho$ is the number of free parameters.

The simplest convergence criterion is based on lack of progress in the log-likelihood, where the algorithm is terminated when $l^{(t+1)}-l^{(t)}<\epsilon$, where $\epsilon>0$ is a small number. Oftentimes, however, the log-likelihood can plateau and then increase again, thus the algorithm would be terminated prematurely using lack of progress, \citep[see][for examples]{mcnicholas10a}. Another option, and one that is used for our analyses, is a criterion based on the Aitken acceleration \citep{aitken26}. The Aitken acceleration at iteration $t$ is
$$
a^{(t)}=\frac{l^{(t+1)}-l^{(t)}}{l^{(t)}-l^{(t-1)}},
$$
where $l^{(t)}$ is the (observed) log-likelihood at iteration $t$. We then have an estimate, at iteration $t+1$, of the log-likelihood after many iterations:
$$
l_{\infty}^{(t+1)}=l^{(t)}+\frac{(l^{(t+1)}-l^{(t)})}{1-a^{(t)}}
$$
\citep{bohning94,lindsay95}. As suggested by \cite{mcnicholas10a}, the algorithm is terminated when $l_{\infty}^{(k+1)}-l^{(k)}\in(0,\epsilon)$. It should be noted that we set the value of $\epsilon$ based on the magnitude of the log-likelihood in the manner of \cite{gallaugher19c}. Specifically, we set $\epsilon$ to a value three orders of magnitude lower than the log-likelihood after five iterations.

To assess classification performance, the adjusted Rand index \cite[ARI;][]{hubert85} is used. The ARI is the Rand index \citep{rand71} corrected for chance agreement. The ARI compares two different partitions---in our case, predicted and true classifications---and takes a value of 1 if there is perfect agreement. Under random classification, the expected values of the ARI is 0. 

Finally, there is the issue of initialization. In our simulations and data analyses, we used soft (uniform) random initializations for the $\hat{z}_{ig}$. From these initial soft group memberships $\hat{z}_{ig}$, we initialize the location matrices using 
$$
\hat{\matm}_g=\frac{1}{N_g}\sum_{i=1}^N{\hat{z}_{ig}\vecX_i},
$$ where $N_g=\sum_{i=1}^N\hat{z}_{ig}$.
The diagonal scale matrices, $\matsig_g$ and $\matPsi_g$ are initialized as follows
$$
\hat{\matsig}_g=\frac{1}{pN_g}\diag\left\{{\sum_{i=1}^N\hat{z}_{ig}(\vecX_i-\hat{\matm}_g)(\vecX_i-\hat{\matm}_g)'}\right\}
$$
and 
$$
\hat{\matPsi}_g=\frac{1}{nN_g}\diag\left\{{\sum_{i=1}^N\hat{z}_{ig}(\vecX_i-\hat{\matm}_g)'(\vecX_i-\hat{\matm}_g)}\right\}.
$$
The elements of the factor loading matrices are initialized randomly from a uniform distribution on $[-1,1]$. Note that all initializations are based on the UUU model.

\section{Simulations}
\subsection{Simulation 1}
Three simulations were conducted. In the first, we consider $d\times d$ matrices with $d\in\{10, 20\}$, $G=2$ and $\matm_1={\bf 0}, \matm_2=\matm_{\text{LT}}^{(\delta)}$, where $\delta\in\{1,2,4\}$ and $\matm_{\text{LT}}^{(\delta)}$ represents a lower triangular matrix with $\delta$ on and below the diagonal. We consider the case where both rows and columns have a CCU model. The parameters for the column factor loading matrices are: 
$$
\loada_1=\loada_2=
\left[
\begin{array}{ccc}
{\bf 1}_{5}&{\bf 0}_{5}&{\bf 0}_{5}\\
{\bf 0}_{2}&{\bf 1}_{2}&{\bf 0}_{2}\\
{\bf 0}_{3}&{\bf 0}_{3}&{\bf 1}_{3}\\
\end{array}
\right] (d=10),
\qquad
\loada_1=\loada_2=
\left[
\begin{array}{ccc}
{\bf 1}_{10}&{\bf 0}_{10}&{\bf 0}_{10}\\
{\bf 0}_{4}&{\bf 1}_{4}&{\bf 0}_{4}\\
{\bf 0}_{6}&{\bf 0}_{6}&{\bf 1}_{6}\\
\end{array}
\right] (d=20).
$$
The row factor loading matrices are
$$
\loadb_1=\loadb_2=\left[
\begin{array}{cc}
-{\bf 1}_{d/2}&{\bf 0}_{d/2}\\
{\bf 1}_{d/2}&{\bf 1}_{d/2}\\
\end{array}
\right],
$$
where ${\bf 1}_c$ and ${\bf 0}_c$ represent $c$-dimensional vectors of 1s and 0s, respectively. The error covariance matrices are taken to be
$$\matsig_1=\matsig_2=\matPsi_1=\matPsi_2={\bf D},$$
where ${\bf D}$ is a diagonal matrix with diagonal entries $d_{tt}=t/5$ when $d=10$ and $d_{tt}=t/10$ when $d=20$.

Finally, sample sizes of $N\in\{100, 200, 400\}$ are considered with $\pi_1=\pi_2=0.5$. For each of these combinations, 25 datasets are simulated. The model is fit for $G=1,\ldots, 4$ groups, 1 to 5 row factors and column factors, and all 64 scale models, leading to a total of 6,400 models fit for each dataset.

In Table \ref{tab:resq1}, we display the number of times the correct number of groups, row factors, and column factors are selected by the BIC, as well as the number of times the row and column models were correctly identified. We also include the average ARI over the 25 datasets with associated standard deviations. As expected, as the separation and sample size increase, better classification results are obtained. The correct number of groups, column factors, and row factors are chosen for all 25 datasets in nearly all cases considered. Moreover, the selection of the row and column models is very accurate in all cases considered.
\begin{table}[!t]
\caption{Number of datasets for which the BIC correctly chose the number of groups ($G$), column factors ($q$), row factors ($r$), row model (RM), column model (CM), and the average ARI over 25 datasets (Simulation 1)}
\label{tab:resq1}       
%
%
\begin{tabular*}{1.00\textwidth}{@{\extracolsep{\fill}}p{0.4cm}p{1cm}p{0.5cm}p{0.5cm}p{0.5cm}p{0.5cm}p{0.5cm}p{1.8cm}p{0.15cm}p{0.5cm}p{0.5cm}p{0.5cm}p{0.5cm}p{0.5cm}p{1.8cm}}
\hline
&&\multicolumn{6}{c}{$d=10$}&&\multicolumn{6}{c}{$d=20$}\\
\cline{3-8}\cline{10-15}\\[-9pt]
$\delta$&$N$&$G$&$q$&$r$&RM& CM&$\overline{\text{ARI}}$(sd)&&$G$&$q$&$r$&RM&CM&$\overline{\text{ARI}}$(sd)\\
\hline
\multirow{3}{*}{$1$}
&$100$&0&25&25&25&25&0.000(0.00)&&25&24&25&25&25&1.000(0.00)\\
&$200$&21&25&25&25&25&0.723(0.33)&&25&24&24&25&24&1.000(0.00)\\
&$400$&25&25&25&25&25&0.883(0.04)&&25&25&25&25&25&1.000(0.002)\\
\hline
\multirow{3}{*}{$2$}
&$100$&25&25&25&25&25&1.000(0.00)&&25&24&25&25&25&1.000(0.00)\\
&$200$&25&25&25&25&25&0.999(0.004)&&25&25&25&25&25&1.000(0.00)\\
&$400$&25&25&25&25&25&1.000(0.002)&&25&25&25&25&25&1.000(0.00)\\
\hline
\multirow{3}{*}{$4$}
&$100$&25&25&25&25&25&1.000(0.00)&&25&24&25&25&25&1.000(0.00)\\
&$200$&25&25&24&25&25&1.000(0.00)&&25&25&25&25&25&1.000(0.00)\\
&$400$&25&25&25&25&25&1.000(0.00)&&25&25&25&25&25&1.000(0.00)\\
\hline
\end{tabular*}
\end{table}

\subsection{Simulation 2}
In this simulation, similar conditions to Simulation 1 are considered, including using the same mean matrices; however, we place a CUC model on the rows and a UCU model on the columns. The column factor loading matrices are the same as used for Simulation 1, $\loadb_1$ is the same as in Simulation 1, and the row factor loadings matrix for group 2 is
$$\loadb_2=\left[
\begin{array}{cc}
{\bf 1}_{d/2}&-{\bf 1}_{d/2}\\
{\bf 1}_{d/2}&{\bf 0}_{d/2}\\
\end{array}
\right].$$
We take $\matsig_1=\ident_d, \matsig_2=2\ident_d$ and $\matPsi_1=\matPsi_2={\bf D}$, where ${\bf D}$ is the same as from Simulation 1.

Results are displayed in Table \ref{tab:resq2}. Overall, we obtain excellent classification results, even when the sample size is small and there is little spatial separation. There is some difficulty in choosing the column model when $d=10$ but this issue abates for $N=400$. When $d=20$, some difficulty is encountered in choosing the correct number of column factors $q$; however, the classification performance is consistently excellent.
\begin{table}[!t]
\caption{Number of datasets for which the BIC correctly chose the number of groups ($G$), column factors ($q$), row factors ($r$), row model (RM), column model (CM), and the average ARI over 25 datasets (Simulation 2)}
\label{tab:resq2}       
%
%
\begin{tabular*}{1.00\textwidth}{@{\extracolsep{\fill}}p{0.4cm}p{1cm}p{0.5cm}p{0.5cm}p{0.5cm}p{0.5cm}p{0.5cm}p{1.8cm}p{0.15cm}p{0.5cm}p{0.5cm}p{0.5cm}p{0.5cm}p{0.5cm}p{1.8cm}}
\hline
&&\multicolumn{6}{c}{$d=10$}&&\multicolumn{6}{c}{$d=20$}\\
\cline{3-8}\cline{10-15}\\[-9pt]
$\delta$&$N$&$G$&$q$&$r$&RM& CM&$\overline{\text{ARI}}$(sd)&&$G$&$q$&$r$&RM&CM&$\overline{\text{ARI}}$(sd)\\
\hline
\multirow{3}{*}{$1$}
&$100$&25&25&25&25&25&0.990(0.02)&&25&0&25&25&25&1.000(0.00)\\
&$200$&25&25&25&25&1&0.998(0.007)&&25&24&25&25&25&1.000(0.00)\\
&$400$&25&25&25&25&25&0.997(0.006)&&25&25&25&25&25&1.000(0.00)\\
\hline
\multirow{3}{*}{$2$}
&$100$&25&25&25&25&0&0.998(0.01)&&25&0&25&25&25&1.000(0.00)\\
&$200$&25&25&25&25&0&1.000(0.00)&&25&24&25&25&25&1.000(0.00)\\
&$400$&25&25&25&25&25&0.999(0.003)&&25&25&25&25&25&1.000(0.00)\\
\hline
\multirow{3}{*}{$4$}
&$100$&25&25&25&25&0&1.000(0.00)&&25&10&25&25&25&1.000(0.00)\\
&$200$&25&25&25&25&2&1.000(0.00)&&25&23&25&25&25&1.000(0.00)\\
&$400$&25&25&24&25&25&1.000(0.00)&&25&5&25&25&20&1.000(0.00)\\

\hline
\end{tabular*}
\end{table}

\subsection{Simulation 3}
In the last simulation, the mean matrices are now diagonal with diagonal entries equal to $\delta$. A CCU model is taken for the rows. In the case of $d=10$, the parameters are 
$$
\loada_1=\loada_2=
\left[
\begin{array}{ccc}
{\bf 1}_3&{\bf 0}_3&{\bf 0}_3\\
{\bf 1}_2&{\bf 0}_2&{\bf 1}_2\\
-{\bf 1}_2&-{\bf 1}_2&-{\bf 1}_2\\
-{\bf 1}_3&-{\bf 1}_3&{\bf 0}_3\\
\end{array}
\right],
\qquad
\matsig_1=\matsig_2=\ident_{d\{\sigma_{2,2}=2, \sigma_{9,9}=4\}}.
$$
To clarify this notation, the row scale matrices have 1s on the diagonal except for places 2 and 9 which have values 2 and 4 respectively. The column scale matrices have a UCC model with 
$$
\loadb_1=
\left[
\begin{array}{cc}
-{\bf 1}_5&{\bf 0}_5\\
{\bf 1}_5&{\bf 1}_5\\
\end{array}
\right],
\qquad
\loadb_2=
\left[
\begin{array}{cc}
-{\bf 1}_5&{\bf 1}_5\\
{\bf 1}_5&{\bf 0}_5\\
\end{array}
\right],
\qquad
\matPsi_1=\matPsi_2=\ident_{10}.
$$
In the case of $d=20$, the parameters are
$$
\loada_1=\loada_2=
\left[
\begin{array}{ccc}
{\bf 1}_{6}&{\bf 0}_{6}&{\bf 0}_{6}\\
{\bf 1}_4&{\bf 0}_4&{\bf 1}_4\\
-{\bf 1}_4&-{\bf 1}_4&-{\bf 1}_4\\
-{\bf 1}_{6}&-{\bf 1}_{6}&{\bf 0}_{6}\\
\end{array}
\right],
\qquad
\matsig_1=\matsig_2=\ident_{30\{\sigma_{2,2}=4, \sigma_{9,9}=2, \sigma_{12,12}=3, \sigma_{19,19}=5\}},
$$
and 
$$
\loadb_1=
\left[
\begin{array}{cc}
-{\bf 1}_{10}&{\bf 0}_{10}\\
{\bf 1}_{10}&{\bf 1}_{10}\\
\end{array}
\right],
\qquad
\loadb_2=
\left[
\begin{array}{cc}
-{\bf 1}_{10}&{\bf 1}_{10}\\
{\bf 1}_{10}&{\bf 0}_{10}\\
\end{array}
\right],
\qquad
\matPsi_1=\matPsi_2=\ident_{20}.
$$
The results are presented in \tablename~\ref{tab:resq3}. In this case, there is more variability in the correct selection of the row and column models, especially the latter. The selection of $q$ and $r$ is generally accurate. The classification performance is generally very good with the exception of the combination of a small sample size $N$ with a low degree of separation $\delta$.

\begin{table}[!t]
\caption{Number of datasets for which the BIC correctly chose the number of groups ($G$), column factors ($q$), row factors ($r$), row model (RM), column model (CM), and the average ARI over 25 datasets (Simulation 3)}
\label{tab:resq3}       
%
%
\begin{tabular*}{1.00\textwidth}{@{\extracolsep{\fill}}p{0.4cm}p{1cm}p{0.5cm}p{0.5cm}p{0.5cm}p{0.5cm}p{0.5cm}p{1.8cm}p{0.15cm}p{0.5cm}p{0.5cm}p{0.5cm}p{0.5cm}p{0.5cm}p{1.8cm}}
\hline
&&\multicolumn{6}{c}{$d=10$}&&\multicolumn{6}{c}{$d=20$}\\
\cline{3-8}\cline{10-15}\\[-9pt]
$\delta$&$N$&$G$&$q$&$r$&RM& CM&$\overline{\text{ARI}}$(sd)&&$G$&$q$&$r$&RM&CM&$\overline{\text{ARI}}$(sd)\\
\hline
\multirow{3}{*}{$1$}
&$100$&0&25&25&25&0&0.000(0.00)&&0&25&25&25&0&0.000(0.00)\\
&$200$&0&25&25&20&0&0.000(0.00)&&0&25&25&25&0&0.000(0.00)\\
&$400$&22&12&25&12&16&0.705(0.27)&&22&21&24&19&13&0.833(0.32)\\
\hline
\multirow{3}{*}{$2$}
&$100$&25&24&25&24&17&0.968(0.04)&&21&24&25&25&10&0.840(0.37)\\
&$200$&25&25&25&25&11&0.984(0.02)&&25&25&25&25&11&1.000(0.00)\\
&$400$&25&20&25&18&22&0.988(0.01)&&25&25&25&25&20&1.000(0.00)\\
\hline
\multirow{3}{*}{$4$}
&$100$&25&24&25&24&15&1.000(0.00)&&25&25&25&25&18&1.000(0.00)\\
&$200$&25&25&25&25&10&1.000(0.00)&&25&25&25&25&22&1.000(0.00)\\
&$400$&25&24&25&20&23&1.000(0.00)&&25&25&25&25&17&1.000(0.00)\\

\hline
\end{tabular*}
\end{table}

\section{MNIST Data Analysis}
\cite{gallaugher18a,gallaugher18b} consider the MNIST digits dataset; specifically, digits~1 and~7 because they are quite similar in appearance. Herein, we consider digits~1 and~2. This dataset consists of 60,000 (training) images of Arabic numerals 0 to 9. We consider different levels of supervision and perform either clustering or semi-supervised classification. Specifically we look at 0\% (clustering), 25\%, and 50\% supervision. For each level of supervision, 25 datasets consisting of 200 images each of digits 1 and 2 are taken. As discussed in \cite{gallaugher18a}, because of the lack of variability in the outlying rows and columns of the data matrices, random noise is added to ensure non-singularity of the scale matrices. In Table \ref{tab:ARIMNIST}, we present the average ARIs and misclassification rates along with respective standard deviations.
\begin{table}[!t]
\caption{Average ARI values and misclassification rates for each level of supervision, with respective standard deviations in parentheses, for datasets consisting of digits 1 and 2 drawn from the MNIST dataset}
\label{tab:ARIMNIST}       
%
%
\begin{tabular*}{1.00\textwidth}{@{\extracolsep{\fill}}p{3.8cm}p{3.8cm}r}
\hline
Supervision&ARI&Misclassification rate\\
\hline
0\% (clustering)&0.652(0.05)&0.0962(0.02) \\
25\%&0.733(0.059)&0.072(0.02) \\
50\% &0.756(0.064)&0.065(0.018) \\
\hline
\end{tabular*}
\end{table}

\begin{figure}[!tb]
\centering
\includegraphics[width=0.675\textwidth]{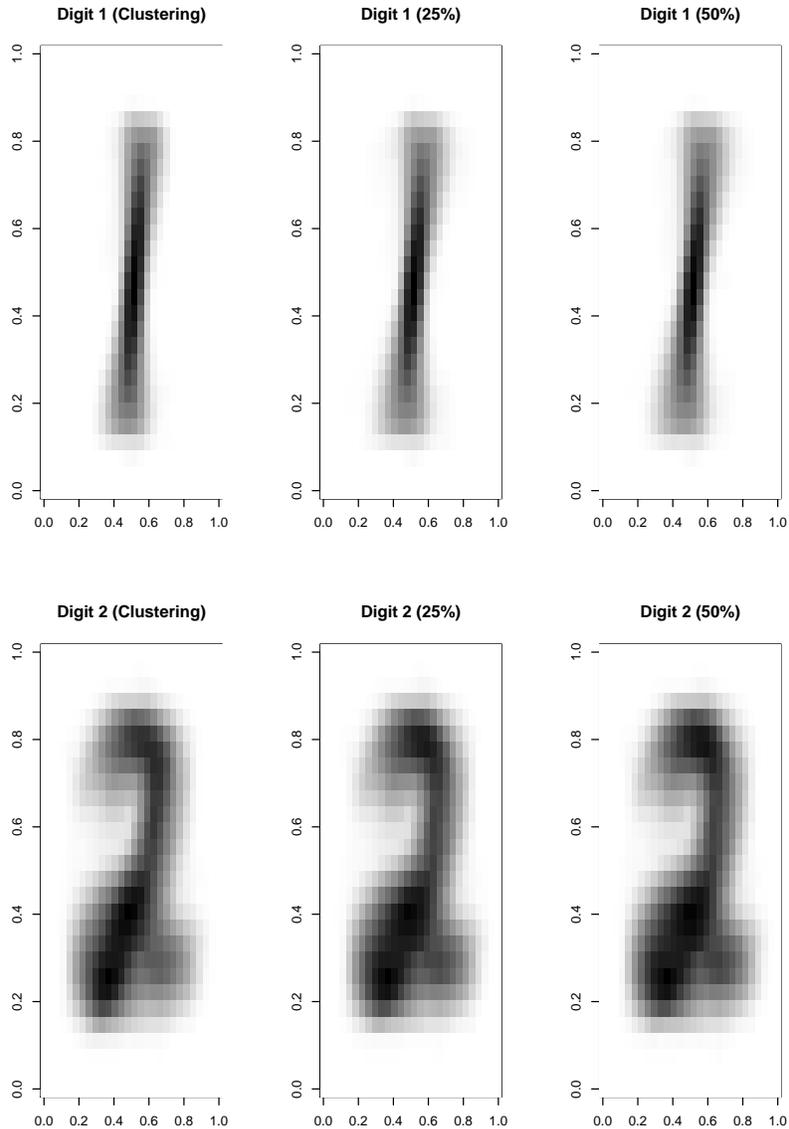}
\caption{Heatmaps of the mean matrices, from one of the datasets, for each digit at each level of supervision.}
\label{fig:heatmap}
\end{figure}

As expected, as the level of supervision is increased, better classification performance is obtained. Specifically, the MCR decreases to around 6.5\% with an ARI of 0.756 when the level of supervision is raised to 50\%. Moreover, the performance in the completely unsupervised case is fairly good. In Figure~\ref{fig:heatmap}, heatmaps for the estimated mean matrices, for one dataset, for each digit and level of supervision are presented. Although barely perceptible, there is a slight increase in clarity as the supervision is raised to 50\%.  For all levels of supervision, the UUU row model is chosen for all 25 datasets. The chosen model for the columns is the UCU model for 7 of the 25 datasets for 0\% and 50\% supervision, and 10 datasets for 25\% supervision. 

\section{Discussion}
The PMMVBFA family, which comprises a total of 64 models has been introduced presented. The PMMVBFA family is essentially a matrix variate analogue of the family of multivariate models introduced by \cite{mcnicholas08}. In all simulations considered, very good classification results were obtained. In some settings, this was true even for small sample sizes and a small amount of spatial separation. Model selection was also generally accurate.
In the MNIST analysis, classification accuracy increased with the amount of supervision, as expected, with the MCR decreasing to 6.5\% when using a supervision level of 50\%. 

One very important consideration for future work is the development of a search algorithm as an alternative to fitting all possible models. This would be particularly important for computational feasibility when there is limited or parallelization available. These parsimonious models could also be extended to the mixtures of skewed matrix variate bilinear factor analyzers \citep{gallaugher19c} as well as an extension to multi-way data \citep[e.g., in the fashion of][]{tait19}.


\appendix
\section{Updates for Scale Matrices and Factor Loadings}
The updates for the scale matrices and the factor loading matrices in the AECM algorithm are dependent on the model. The exact updates for each model are presented here.

\subsection{Row Model Updates}

\textbf{CCC}:
\begin{equation*}\begin{split}
\hat{\loada}&=\left(\sum_{g=1}^G\sum_{i=1}^N\hat{z}_{ig}(\vecX_i-\hat{\matm}_g)\hat{\matPsi}^{*^{-1}}_g{a_{ig}^A}'\right)\left(\sum_{g=1}^G\sum_{i=1}^N \hat{z}_{ig}b_{ig}^B\right)^{-1},\ \
\hat{\sigma}=\frac{1}{Nnp}\tr\{\vecS^{(1)}\}.
\end{split}\end{equation*}
where 
$$
\vecS^{(1)}=\sum_{g=1}^G\sum_{i=1}^N\hat{z}_{ig}\big[(\vecX_i-\hat{\matm}_g)\hat{\matPsi}^{*^{-1}}_g(\vecX_i-\hat{\matm}_g)'-\hat{\loada}{a_{ig}^B}'\hat{\matPsi}^{*^{-1}}_g(\vecX_i-\hat{\matm}_g)'\big].
$$

\noindent \textbf{CCU}:
\begin{equation*}\begin{split}
\hat{\loada}&=\left(\sum_{g=1}^G\sum_{i=1}^N\hat{z}_{ig}(\vecX_i-\hat{\matm}_g)\hat{\matPsi}^{*^{-1}}_g{a_{ig}^B}'\right)\left(\sum_{g=1}^G\sum_{i=1}^N \hat{z}_{ig}b_{ig}^B\right)^{-1},\ \
\hat{\matsig}=\frac{1}{Np}\diag\{\vecS^{(1)}\},
\end{split}\end{equation*}

\noindent \textbf{CUU}:\\
For this model, the update for $\loada$ needs to be performed row by row. Specifically, the updates are:
\begin{equation*}\begin{split}
\hat{\loada}_{(j)}&=\left(\sum_{i=1}^N\hat{z}_{ig}(\vecX_i-\hat{\matm}_g)\hat{\matPsi}^{*^{-1}}_g{a_{ig}^B}'\right)_{(j)}\left(\sum_{g=1}^G\frac{1}{\sigma_{{g}_{(jj)}}}\sum_{i=1}^N \hat{z}_{ig}b_{ig}^B\right)^{-1},\\
\hat{\matsig}_g&=\frac{1}{N_gp}\diag\{\vecS_g^{(2)}\},
\end{split}\end{equation*}
where 
$$\vecS_g^{(2)}=\sum_{i=1}^N\hat{z}_{ig}\big[(\vecX_i-\hat{\matm}_g)\hat{\matPsi}^{*^{-1}}_g(\vecX_i-\hat{\matm}_g)'-2\hat{\loada}{a_{ig}^B}\hat{\matPsi}^{*^{-1}}_g(\vecX_i-\hat{\matm}_g)'+\hat{\loada}b_{ig}^B\hat{\loada}'\big].$$

\noindent \textbf{CUC}:
\begin{equation*}\begin{split}
\hat{\loada}&=\left(\sum_{g=1}^G\frac{1}{\hat{\sigma}_g}\sum_{i=1}^N\hat{z}_{ig}(\vecX_i-\hat{\matm}_g)\hat{\matPsi}^{*^{-1}}_g{a_{ig}^B}'\right)\left(\sum_{g=1}^G\frac{1}{\hat{\sigma}_g}\sum_{i=1}^N \hat{z}_{ig}b_{ig}^B\right)^{-1},\\
\hat{\sigma}_g&=\frac{1}{N_gnp}\tr\{\vecS_g^{(2)}\}.
\end{split}\end{equation*}

\noindent \textbf{UCC}:
\begin{equation*}\begin{split}
\hat{\loada}_g&=\left(\sum_{i=1}^N\hat{z}_{ig}(\vecX_i-\hat{\matm}_g)\hat{\matPsi}^{*^{-1}}_g{a_{ig}^B}'\right)\left(\sum_{i=1}^N \hat{z}_{ig}b_{ig}^B\right)^{-1},\ \
\hat{\sigma}=\frac{1}{Nnp}\tr\{\vecS^{(3)}\},
\end{split}\end{equation*}
where 
$$\vecS^{(3)}=\sum_{g=1}^G\sum_{i=1}^N\hat{z}_{ig}\big[(\vecX_i-\hat{\matm}_g)\hat{\matPsi}^{*^{-1}}_g(\vecX_i-\hat{\matm}_g)'-\hat{\loada}_g{a_{ig}^B}'\hat{\matPsi}^{*^{-1}}_g(\vecX_i-\hat{\matm}_g)'\big].$$

\noindent \textbf{UCU}:
\begin{equation*}\begin{split}
\hat{\loada}_g&=\left(\sum_{i=1}^N\hat{z}_{ig}(\vecX_i-\hat{\matm}_g)\hat{\matPsi}^{*^{-1}}_g{a_{ig}^B}'\right)\left(\sum_{i=1}^N \hat{z}_{ig}b_{ig}^B\right)^{-1},\ \
\hat{\matsig}=\frac{1}{Np}\diag\{\vecS^{(3)}\}.
\end{split}\end{equation*}

\noindent \textbf{UUC}:

\begin{equation*}\begin{split}
\hat{\loada}_g&=\left(\sum_{i=1}^N\hat{z}_{ig}(\vecX_i-\hat{\matm}_g)\hat{\matPsi}^{*^{-1}}_g{a_{ig}^B}'\right)\left(\sum_{i=1}^N \hat{z}_{ig}b_{ig}^B\right)^{-1},\ \
\hat{\sigma}_g=\frac{1}{N_gnp}\tr\{\vecS_g^{(4)}\},
\end{split}\end{equation*}
where 
$$\vecS_g^{(4)}=\sum_{i=1}^N\hat{z}_{ig}\big[(\vecX_i-\hat{\matm}_g)\hat{\matPsi}^{*^{-1}}_g(\vecX_i-\hat{\matm}_g)'-\hat{\loada}_g{a_{ig}^B}\hat{\matPsi}^{*^{-1}}_g(\vecX_i-\hat{\matm}_g)'\big].$$

\noindent \textbf{UUU}:
\begin{equation*}\begin{split}
\hat{\loada}_g&=\left(\sum_{i=1}^N\hat{z}_{ig}(\vecX_i-\hat{\matm}_g)\hat{\matPsi}^{*^{-1}}_g{a_{ig}^B}'\right)\left(\sum_{i=1}^N \hat{z}_{ig}b_{ig}^B\right)^{-1},\ \
\hat{\matsig}_g=\frac{1}{N_gp}\diag\{\vecS_g^{(4)}\}.
\end{split}\end{equation*}

\subsection{Column Model Updates}

\noindent \textbf{CCC}:
\begin{equation*}\begin{split}
\hat{\loadb}&=\left(\sum_{g=1}^G\sum_{i=1}^N\hat{z}_{ig}(\vecX_i-\hat{\matm}_g)'\hat{\matsig}^{*^{-1}}_g{a_{ig}^A}\right)\left(\sum_{g=1}^G\sum_{i=1}^N \hat{z}_{ig}b_{ig}^A\right)^{-1},\ \
\hat{\psi}=\frac{1}{Nnp}\tr\{\vecP^{(1)}\},
\end{split}\end{equation*}
where 
$$
\vecP^{(1)}=\sum_{g=1}^G\sum_{i=1}^N\hat{z}_{ig}\big[(\vecX_i-\hat{\matm}_g)'\hat{\matsig}^{*^{-1}}_g(\vecX_i-\hat{\matm}_g)-\hat{\loadb}{a_{ig}^A}'\hat{\matsig}^{*^{-1}}_g(\vecX_i-\hat{\matm}_g)\big].
$$

\noindent \textbf{CCU}:
\begin{equation*}\begin{split}
\hat{\loadb}&=\left(\sum_{g=1}^G\sum_{i=1}^N\hat{z}_{ig}(\vecX_i-\hat{\matm}_g)'\hat{\matsig}^{*^{-1}}_g{a_{ig}^A}\right)\left(\sum_{g=1}^G\sum_{i=1}^N \hat{z}_{ig}b_{ig}^A\right)^{-1},\ \
\hat{\matPsi}=\frac{1}{Nn}\diag\{\vecP^{(1)}\}.
\end{split}\end{equation*}

\noindent \textbf{CUU}:\\
For this model, the update for $\loadb$ needs to be performed row by row.
Specifically the updates are:
\begin{equation*}\begin{split}
\hat{\loadb}_{(j)}&=\left(\sum_{i=1}^N\hat{z}_{ig}(\vecX_i-\hat{\matm}_g)'\hat{\matsig}^{*^{-1}}_g{a_{ig}^A}\right)_{(j)}\left(\sum_{g=1}^G\frac{1}{\psi_{{g}_{(jj)}}}\sum_{i=1}^N \hat{z}_{ig}b_{ig}^A\right)^{-1},\\
\hat{\matPsi}_g&=\frac{1}{N_gn}\diag\{\vecP_g^{(2)}\},
\end{split}\end{equation*}
where 
$$
\vecP_g^{(2)}=\sum_{i=1}^N\hat{z}_{ig}\big[(\vecX_i-\hat{\matm}_g)'\hat{\matsig}^{*^{-1}}_g(\vecX_i-\hat{\matm}_g)-2\hat{\loadb}{a_{ig}^A}'\hat{\matsig}^{*^{-1}}_g(\vecX_i-\hat{\matm}_g)+\hat{\loadb}b_{ig}^A\hat{\loadb}'\big].
$$

\noindent \textbf{CUC}:
\begin{equation*}\begin{split}
\hat{\loadb}&=\left(\sum_{g=1}^G\frac{1}{\hat{\psi}_g}\sum_{i=1}^N\hat{z}_{ig}(\vecX_i-\hat{\matm}_g)'\hat{\matsig}^{*^{-1}}_g{a_{ig}^A}\right)\left(\sum_{g=1}^G\frac{1}{\hat{\psi}_g}\sum_{i=1}^N \hat{z}_{ig}b_{ig}^A\right)^{-1},\\
\hat{\psi}_g&=\frac{1}{N_gnp}\tr\{\vecP_g^{(2)}\}.
\end{split}\end{equation*}

\noindent \textbf{UCC}:
\begin{equation*}\begin{split}
\hat{\loadb}_g&=\left(\sum_{i=1}^N\hat{z}_{ig}(\vecX_i-\hat{\matm}_g)'\hat{\matsig}^{*^{-1}}_g{a_{ig}^A}\right)\left(\sum_{i=1}^N \hat{z}_{ig}b_{ig}^A\right)^{-1},\ \
\hat{\psi}=\frac{1}{Nnp}\tr\{\vecP^{(3)}\},
\end{split}\end{equation*}
where 
$$
\vecP^{(3)}=\sum_{g=1}^G\sum_{i=1}^N\hat{z}_{ig}\big[(\vecX_i-\hat{\matm}_g)'\hat{\matsig}^{*^{-1}}_g(\vecX_i-\hat{\matm}_g)-\hat{\loadb}_g{a_{ig}^A}'\hat{\matsig}^{*^{-1}}_g(\vecX_i-\hat{\matm}_g)\big].
$$

\noindent \textbf{UCU}:
\begin{equation*}\begin{split}
\hat{\loadb}_g&=\left(\sum_{i=1}^N\hat{z}_{ig}(\vecX_i-\hat{\matm}_g)'\hat{\matsig}^{*^{-1}}_g{a_{ig}^A}\right)\left(\sum_{i=1}^N \hat{z}_{ig}b_{ig}^A\right)^{-1},\ \
\hat{\matPsi}=\frac{1}{Nn}\diag\{\vecP^{(3)}\}.
\end{split}\end{equation*}

\noindent \textbf{UUC}:
\begin{equation*}\begin{split}
\hat{\loadb}_g&=\left(\sum_{i=1}^N\hat{z}_{ig}(\vecX_i-\hat{\matm}_g)'\hat{\matsig}^{*^{-1}}_g{a_{ig}^A}\right)\left(\sum_{i=1}^N \hat{z}_{ig}b_{ig}^A\right)^{-1},\ \
\hat{\psi}_g=\frac{1}{N_gnp}\tr\{\vecP_g^{(4)}\},
\end{split}\end{equation*}
where 
$$\vecP_g^{(4)}=\sum_{i=1}^N\hat{z}_{ig}\big[(\vecX_i-\hat{\matm}_g)'\hat{\matsig}^{*^{-1}}_g(\vecX_i-\hat{\matm}_g)-\hat{\loadb}_g{a_{ig}^A}'\hat{\matsig}^{*^{-1}}_g(\vecX_i-\hat{\matm}_g)\big].$$

\noindent \textbf{UUU}:
\begin{equation*}\begin{split}
\hat{\loadb}_g&=\left(\sum_{i=1}^N\hat{z}_{ig}(\vecX_i-\hat{\matm}_g)'\hat{\matsig}^{*^{-1}}_g{a_{ig}^A}\right)\left(\sum_{i=1}^N \hat{z}_{ig}b_{ig}^A\right)^{-1}, \ \
\hat{\matPsi}_g=\frac{1}{N_gn}\diag\{\vecP_g^{(4)}\}.
\end{split}\end{equation*}

\end{document}